# Why Mathematics Works So Well

Noson S. Yanofsky

A major question in philosophy of science involves the unreasonable effectiveness of mathematics in physics. Why should mathematics, created or discovered, with nothing empirical in mind be so perfectly suited to describe the laws of the physical universe? We review the well-known fact that the symmetries of the laws of physics are their defining properties. We show that there are similar symmetries of mathematical facts and that these symmetries are the defining properties of mathematics. By examining the symmetries of physics and mathematics, we show that the effectiveness is actually quite reasonable. In essence, we show that the regularities of physics are a subset of the regularities of mathematics.

**Introduction**.

One of the most interesting problems in philosophy of science and philosophy of mathematics is concerned with the relationship between the laws of physics and the world of mathematics. Why should mathematics so perfectly describe the workings of the universe? Significant areas of mathematics are formed without anything physical in mind, and yet such mathematics can be used to describe the laws of physics. How are we to understand this?

This mystery is seen most clearly by examining the power of mathematics to determine the existence of physical objects before there is empirical evidence of those objects. One of the more famous examples of the predictive abilities of mathematics is the discovery of Neptune by Urbain Le Verrier simply by making some calculation about the abnormalities of the orbit of Uranus. Other examples are P.A.M. Dirac's prediction of the existence of positrons and James Clerk Maxwell's extrapolation that varying electric or magnetic fields should generate propagating waves.

Even more amazing is that there existed entire established fields of mathematics long before physicists realized that they were useful for understanding various aspects of the physical universe. The conic sections studied by Apollonius in ancient Greece were used by Johannes Kepler in the beginning of the seventeenth century to understand the orbits of the planets. Complex numbers were invented several centuries before physicists started using them to describe quantum mechanics. Non-Euclidian geometry was developed decades before it was used in an essential way for general relativity. (Details of these and other remarkable mathematical discoveries can be found in [Yan].)

Why should mathematics be so good at describing the world? Of all thoughts, ideas, or ways of expressing things, why should mathematics work so well? What about other modes of thought? Why does poetry fail to describe the exact movements of the celestial bodies? Why can't music capture the full complexity of the periodic table? Why is meditation not very helpful in predicting the outcomes of experiments in quantum mechanics?

The problem of why mathematics works so well was famously addressed by Nobel prize winning physicist Eugene Wigner in a paper titled "The unreasonable effectiveness of mathematics in the natural sciences" [Wigner1]. Wigner did not arrive at any definitive answers to the questions. He wrote that "the enormous usefulness of mathematics in the natural sciences is something bordering on the mysterious and … there is no rational explanation for it."

Albert Einstein perfectly described the mystery as follows:

> How can it be that mathematics, being after all a product of human thought which is independent of experience, is so admirably appropriate to the objects of reality? Is human reason, then, without experience, merely by taking thought, able to fathom the properties of real things? [Einstein]

To be clear, the problem really arises when one considers both physics and mathematics to each be perfectly formed, objective and independent of human observers. With such a conception, one can ask why these two independent disciplines harmonize so well. Why is it that that an independently discovered law of physics can be described perfectly by (already discovered) mathematics?

Many researchers have pondered this mystery and offered various solutions to the problem. Theologians solve the mystery by positing a Being who perfectly set up the laws of the universe and used the language of mathematics to describe these laws. However the existence of such a Being only adds to the mystery of the universe. Platonists (and their cousins Realists) believe that there exists some realm of "perfect Forms" which contains all mathematical objects, structures and truths. In addition, this "Platonic attic" also contains the physical laws. The problem with Platonism is that, in order to explain the relationship between our mathematical world and the physical universe, it invokes yet another Platonic world. Now one must explain the relationship between all three of these worlds. Other questions also arise: are imperfect mathematical theorems also a perfect Form? Do outdated laws of physics also reside in Plato's attic? Who set up this world of perfect Forms?

The reason most cited (see [Mickens]) to answer the unreasonable effectiveness is that we learn mathematics by examining the physical universe. We understood some of the properties of addition and multiplication by counting stones and sheep. We learned geometry by looking at physical shapes. From this point of view, it is not a surprise that physics follows mathematics since the mathematics that we know was formed by scrutinizing the physical world. The main problem with this solution is that mathematics is very useful in areas distant from human perception. Why is the hidden world of subatomic particles so perfectly described by a mathematics learned by observing stones and sheep? Why is special relativity, which deals with objects that move near the speed of light, described by a mathematics that was learned by watching objects that move at normal speeds?

While these and other purported solutions to the unreasonable effectiveness of mathematics have some merit, the mystery still remains. In two recent papers [YanZel1,YanZel2], Mark Zelcer and I have formulated a novel view of the nature of mathematics. We show that just as symmetry plays an important role in physics, so too, symmetry plays a major role in mathematics. This view of mathematics supplies an original solution to the unreasonable effectiveness problem.

**What is physics?**

Before we can contemplate the reason why mathematics describes physical laws so well, we have to be exact as to the definition of a physical law. To say that a physical law describes a physical phenomenon is a bit naïve. It is much more. As a first attempt we might say that each law describes many different physical phenomena. For example, the law of gravity describes what happens when I drop my spoon; it describes what happens when I drop my spoon tomorrow; and it describes what happens if I drop my spoon on the planet Saturn next month. A law of physics describes a whole bundle of many different phenomena. However this definition is not enough. A single physical phenomenon can be perceived in many different ways. One can perceive a phenomenon while remaining stationary at the same time as another perceives the same phenomenon while moving (in a uniform constant frame, or an accelerating frame). Physics states that no matter how a single phenomenon is perceived, it should be described by a single physical law. We conclude that a physical law describes a whole bundle of *perceived* physical phenomena. For example, the law of gravity describes my observation of a spoon falling in a speeding car while I am in the car; a stationary friend's observation of a spoon falling in a speeding car; an observation by someone standing on his head near a black hole of a falling spoon in a speeding car on Saturn, etc. Every physical law describes a large bundle of perceived physical phenomena.

The question then arises as how to classify all perceived physical phenomena into different laws or which perceived physical phenomena should be bound up together. Physicists use the notion of symmetry for this. Colloquially the word "symmetry" is used to describe physical objects. We say a room has symmetry if the left side of the room is the same as the right side of the room. In other words, if we swap the furnishings from one side to the other, the room would look the same. Scientists have extended this definition of symmetry to describe physical laws. A physical law is symmetric with respect to a type of translation if the law still describes the translated phenomenon. For example, physical laws are symmetric with respect to location. This means that if an experiment is done in Pisa or Princeton, the results of the experiment are the same. The phenomenon occurring in Pisa is bundled up with the phenomenon occurring in Princeton. Physical laws are also symmetric with respect to time, i.e., performing the same experiment today or tomorrow should give us the same result. In terms of bundles, that means that all experiments performed at any time are bundled up to be in the same class of perceived physical phenomena. Another obvious symmetry is orientation. If you change the orientation of an experiment, the results of the experiment remain the same. The languages of bundles of perceived physical phenomena and of symmetries of physical laws are equivalent. In this paper we employ both languages.

There are many other types of symmetries that physical laws have to obey. Galilean relativity demands that the laws of motion remain unchanged if a phenomenon is observed while stationary or moving at a uniform, constant velocity. Special relativity states that the laws of motion must remain the same even if the observers are moving close to the speed of light. General relativity states that the laws are invariant even if the observer is moving in an accelerating frame.

Physicists have generalized the notion of symmetry in many different ways: gauge transformations, local symmetries, global symmetries, continuous symmetries, discrete symmetries, etc. Victor Stenger [Stenger] unites the many different types of symmetries under what he calls *point of view invariance*. That is, all the laws of physics must remain the same regardless of how they are viewed. He demonstrates how much ---but not all--- of modern physics can be recast as laws that satisfy point of view invariance. This means that different perceived physical phenomena are bundled together if they are related to the same physical phenomenon but are perceived from different points of view.

The real importance of symmetry came when Einstein formulated the laws of special relativity. Prior to him, one first found a law of nature and then found its symmetries. In contrast, Einstein used the symmetries to discover the laws. In order to find the laws of special relativity, he posited that the laws must be the same for a stationary observer and an observer moving close to speed of light. Given this presupposition, he went on to formulate the equations that describe special relativity. This was revolutionary. Einstein had realized that symmetries are the defining characteristics of laws of physics. It is not that physical laws satisfy symmetries. Rather, whatever satisfies symmetries *is* a physical law.

In 1918, Emmy Noether showed that symmetry is even more central to physics. She proved a celebrated theorem that connected symmetry to conservation laws that permeate physics. The theorem states that for every symmetry of a special type there exists a conservation law and vice versa. For example, the fact that the laws of physics are invariant with respect to translations in space corresponds to conservation of linear momentum. Time invariance corresponds to conservation of energy. Orientation invariance corresponds to conservation of angular momentum. Equipped with the understanding given by Einstein and Noether of the centrality of symmetry, physicists have been searching for novel and different types of symmetries in order to find new laws of physics.

With this broad definition of physical law, it is not hard to see why the laws have a feeling of being objective, timeless and independent of human observations. Since the laws are applied in every place, at every time, and from every perspective, they have a feeling of being "out there." However one need not look at it that way. Rather than saying that we are looking at many different instances of an external physical law, we may say that we humans select those perceived physical phenomena that have some type of regularity and bundle them together to form a single physical law. We act like a sieve that picks and chooses from all the physical phenomena that we perceive, we bundle together what is the same physical law, and we ignore the rest. We cannot eliminate the human element in understanding the laws of nature.

Before we proceed, we must mention a symmetry that is so obvious it has not been articulated. A law of physics must satisfy *symmetry of applicability*. That is, if a law works for a particular physical object of a certain type then it will work for another physical object of the same type. For example, if a law is true for one positively charged subatomic particle moving at close to the speed of light, then it will work for another positively charged subatomic particle that is also moving at close to the speed of light. In contrast, that law might not work for a macroscopic object moving at a slow speed. All of these different perceived physical phenomena will be bundled together as one law. Symmetry of applicability will be of fundamental importance when we discuss the relationship of physics to mathematics.

**What is mathematics?**

Let us spend a few minutes considering the real essence of mathematics. We illustrate with three examples.

Long ago a farmer realized that if you take nine apples and combine them with four apples, there will be thirteen apples. Not long after that it was noticed that if nine oranges are combined with four oranges, there will be thirteen oranges. That is, if you swap every apple for an orange, the amount of fruit remains the same. At some point an early mathematician looked at many instances of this and bundled them together to summarize with the mathematical expression $9 + 4 = 13$. This pithy little statement encapsulates all the instances of this type of combination. This expression will be true for any whole discrete object that can be exchanged for apples.

On a more sophisticated level, a major theorem in algebraic geometry is *Hilbert's Nullstellensatz* which is essential for understanding the relationship between ideals in polynomial rings and algebraic sets. For every ideal $J$ in a polynomial ring there is a related algebraic set, $V(J)$, and for every algebraic set S there is an ideal I(S). The relationship between these two operations is given as follows: for every ideal $J$ we have $I(V(J)) = \sqrt{J}$ where $\sqrt{J}$ is the radical of the ideal. If we change one ideal for another ideal, we get a different algebraic set. If we swap one algebraic set for another, we get a different ideal. Essentially, the bundling together of all the different instances of this law is Hilbert's theorem.

One of the basic ideas in algebraic topology is the *Hurewicz homomorphism*. For any topological space $X$ and positive integer $k$ there exists a group homomorphism from the $k$-th homotopy group to the $k$-th homology group $h_*: \pi_k(X) \to H_k(X)$. This homomorphism has special properties depending on the space $X$ and the integer $k$. If the space $X$ is swapped for space $Y$ and $k$ is exchanged for $k'$ there will be another group homomorphism $\pi_{k'}(Y) \to H_{k'}(Y)$. Once again, no single instance of the statement has any mathematical content. Rather, it is the realization that all the instances of the statement can be bundled together that makes this mathematics.

In these three examples, we focused on changing the semantics of the mathematical statements. We exchanged oranges for apples. We swapped one ideal for another. We replaced one topological space for a different one. Our main point is that when you make the appropriate changes, the mathematical facts remain true. We claim that this ability to alter the semantics of a mathematical statement is the defining property of mathematics. That is, a statement is mathematics if we can swap what it refers to and remain true.

Associated with every mathematical statement is a class of entities called the *domain of discourse* for that statement. The statement is stating something about all the elements of this domain of discourse. When a mathematician says "For any integer $n$…," "Take a Hausdorff space..," or "Let C be a cocommutative coassociative coalgebra with an involution…," she is setting up the domain of discourse for that statement. If the statement is true for some element in the domain of discourse, it is true for any other. Notice that for a statement the domain of discourse can consist of many types of entities.

This swapping of one element in the domain of discourse for another can be seen as a type of symmetry. If you swap one referent for another referent within the domain of discourse, the fact will remain true. We call this *symmetry of semantics*. Mathematics is invariant with respect to symmetry of semantics. We are claiming that this type of symmetry is as fundamental to mathematics as symmetry is to physics. In the same way that physicists formulate laws, mathematicians formulate mathematical statements by determining which bundle of ideas satisfies symmetry of semantics. We go further than just saying that mathematical truths satisfy symmetry of semantics. Rather, anything that satisfies symmetry of semantics, *is* mathematics.

The logicians among us will find symmetry of semantics to be a familiar notion. They call a logical statement *valid* if it is true for every element of the domain of discourse. Here we are saying that a statement satisfies symmetry of semantics if we can exchange any element of its intended domain of discourse with any other element of its domain of discourse. The novelty we are expressing here is that validity can be seen as a form of symmetry and that this symmetry is the defining feature of mathematics.

One might object to this definition of mathematics as being too broad and say that what we are defining is any general statement. The argument would be that statements that satisfy symmetry of semantics are not just mathematical statements but any general statement. There is a two-prong defense to this criticism. First, modern mathematics is very broad. Mathematics is not just about numbers and quantities. Looking at modern mathematics one finds shapes, propositions, sets, categories, microstates, macrostates, qualities, etc. In order to deal with all these objects, our definition of mathematics needs to be broad. A second defense is that there are many general statements that do not satisfy symmetry of semantics. "It is cold in New York during the month of January," "Flowers are red and green," and "Senators are honest people" are general statements but do not satisfy symmetry of semantics and hence are not mathematical. As long as there are any counterexamples to such statements within their implied domain of discourse, they may be general, but they are not mathematical. The fact that symmetry of semantics does not permit any counterexamples within the domain of discourse implies a certain precision of thought and language which people associate with mathematics.

Mathematical statements also have other types of symmetries that they satisfy. A simple example is *symmetry of syntax*. This says that a mathematical object can be described (syntax) in many different ways. For example we can write 6 as $2 \times 3$, or $2 + 2 + 2$ or $54 \div 9$. Similarly we can talk about a "non-self-intersecting continuous loop," "a simple closed curve," or "a Jordan curve" and mean the same thing. Our point is that the results of the mathematics will be the same regardless of which syntax is used. In practice mathematicians tend to use the simplest syntax possible, like "6" instead of "5+2-1".

Other symmetries that mathematical statements possess are so obvious and taken for granted that even mentioning them seems strange. For example, mathematical truths are invariant with respect to time and space: if they are true now then they will also be true tomorrow. If they are true in Albany they are true on Alpha Centauri. It is similarly immaterial if the mathematical truth is asserted by Mother Teresa or by Oswald Teichmüller. No one cares where, when, or in what language a theorem is stated. It is irrelevant if the mathematical statement was stated at all.

With mathematics satisfying all of these different types of symmetries, it is easy to see why mathematics --like physics -- also has the feel of being objective, timeless, and independent of human observers. Since the facts of mathematics apply to many different objects, are discovered by many different individuals working independently, and in many different times and places, one can start believing that math is somehow "out there." But, we need not make that leap. Symmetry of semantics is at the core of how we determine mathematical truths. Human beings function like sieves that pick and choose from among thoughts and ideas. We bundle the thoughts that are related by symmetry of semantics and declare such statements to be mathematics. We do not say that there exists some perfect mathematical truths and we humans find many different instances of that truth. Rather, we say that there are many different instances of a mathematical fact and we humans bundle them together to form a clear mathematical statement.

See [YanZel1] and [YanZel2] for much more discussion about the nature of mathematics and how other issues in the philosophy of mathematics are dealt with.

**Why mathematics works well at describing physics?**

Armed with this understanding of the nature of physics and mathematics we can tackle the question of why mathematics works so well at expressing physical laws. Let us look at three physical laws.

Our first example is gravity. A description of a single instance of a perceived physical phenomenon of gravity might look like this: "On the second floor of 5775 Main Street in Brooklyn, New York at 9:17:54PM, I saw an 8.46 ounce spoon fall and hit the floor 1.38 seconds later." While it might be totally accurate, it is not very useful and it is not the description of all the instances of the law of gravity. As we explained, a law consists of all the perceived instances of that law. The only way to capture all of the bundled perceptions of physical phenomena of a particular law is to write it in mathematical language which has all its instances bundled with it. Only Newton's formula $F = G \frac{m_1 m_2}{d^2}$ can capture the entire bundle of perceived physical phenomena of gravity. By substituting the mass of one body into $m_1$ and the mass of the other body into $m_2$ and the distance into $d$, we are describing an instance of gravity.

In a similar vein, to find the extremum of an action one needs to use the Euler–Lagrange equations: $\frac{\partial L}{\partial q} = \frac{d}{dt} \frac{\partial L}{\partial \dot{q}}$. The symmetries of an action and its local maximums or minimums can be expressed with these equations that are defined with the symmetry of semantics. Of course, this can also be expressed by a formula that uses other variables or other symbols. By symmetry of syntax, we can even write the formula in Esperanto. It is irrelevant how and in what language the mathematics is expressed. As long as it is mathematics because only mathematics can capture the action principle in all its instances.

The only way to truly encapsulate the relationship between pressure, volume, number of moles and temperature of an ideal gas is the ideal gas law: $PV = nRT$. This small mathematical statement has all the instances of that law built into it.

In the three examples given, all the perceived instances of the physical laws can only be expressed with the mathematical formula. The perceived physical phenomena that we are trying to express are all implied by all the instances which are inherent in the mathematical statement. In terms of symmetry, we are saying that the physical symmetry of applicability is a special type of the mathematical symmetry of semantics. In detail, for any physical law, symmetry of applicability states that the law can deal with swapping any appropriate object for any other appropriate object. If there is a mathematical statement that can describe this physical law, then we can substitute different values for the different objects that one is applying. In terms of bundles, we are saying that every bundle of perceived physical phenomena is a sub-bundle of instances of the mathematical law that describes it.

The point we are making is that mathematics works so well at describing laws of physics because they were both formed in the same way. The laws of physics are not living in some Platonic attic nor are the central ideas of mathematics. Both the physicist and the mathematician chose their statements to be applicable in many different contexts. We bundle perceived physical phenomena in the same way we bundle instances of mathematical truth. It is not a mystery that the abstract laws of physics are stated in the abstract language of mathematics. Rather the regularities of phenomena and thoughts are seen and chosen by human beings in the same way. The fact that some of the mathematics could have been formulated long before the law of physics is discovered is not so strange since they were formed with the same notion of symmetry.

We have not completely solved the mystery concerning the unreasonable effectiveness of mathematics. There are still deep questions lurking here. For one, we can ask why humans even have physics or mathematics. Why do we notice regularities or symmetries? The answer is that part of being alive is being somewhat homeostatic, that is, living creatures must preserve themselves. The more they comprehend their environment, the better off they will be. Inanimate objects like sticks and stones do not react to their environment. In contrast, plants turn towards the sun and their roots search for water. As living creatures become more sophisticated they notice more about their environment. Human beings notice many regularities of the world around them. While chimpanzees do not seem to understand abstract algebra and clever dolphins do not write textbooks in quantum field theory, humans do have the ability to grasp the regularities of their perceived physical phenomena and the thoughts that go through their head. We call the regularities of our thoughts "mathematics" and some of these regularities manifest themselves as regularities of perceived physical phenomena, which we call "physics."

On the deepest level one can ask why there are any regularities at all in perceived physical phenomena? Why should it be that an experiment performed in Poughkeepsie will yield the same results as if it was performed in Piscataway? Why should balls roll down ramps at the same speed even if they are released at different times? Why should chemical reactions be the same when they are perceived by different people and in different ways? To answer such questions, we appeal to the anthropic principle. This is a type of reasoning that formulates answers from the very fact that we exist. If the universe did not have some regularities, no life would be possible. Life uses the fact that the physical universe contains some repeating patterns. Since there is life in the universe, there has to be certain regularities in the laws of physics. If the universe was totally random or like a psychedelic vision, no life ---in particular no intelligent human life--- would survive. Anthropic reasoning does not eliminate the question. It tells us that the fact that we are here means that the universe must be a certain way, and had the universe been another way, we could not exist to ask the question. But the anthropic principle does not tell us *why* the universe is that way. We are still left with those deep ---and, as yet, unanswerable--- questions of "Why is the universe here?" "Why is there something rather than nothing?" and "What's going on here?"

While we have not eliminated all the mysteries, we have shown that any existing structure in our perceived physical universe is naturally expressed in the language of mathematics.